\documentclass{webofc}
\usepackage[varg]{txfonts}   % Web of Conferences font
\def\mkfit{\texttt{mkFit}\xspace}
\def\cmssw{\texttt{CMSSW}\xspace}
\def\tbb{\texttt{TBB}\xspace}
\def\icc{\texttt{icc}\xspace}
\def\gcc{\texttt{gcc}\xspace}
\def\etal{\emph{et al.}\xspace}
\begin{document}
\title{Reconstruction of Charged Particle Tracks in Realistic Detector Geometry Using a Vectorized and Parallelized Kalman Filter Algorithm}

\author{
       \firstname{Giuseppe} \lastname{Cerati}\inst{2} \fnsep\thanks{\email{cerati@fnal.gov}}
  \and \firstname{Peter} \lastname{Elmer}\inst{3} %\fnsep\thanks{\email{peter.elmer@cern.ch}}
  \and \firstname{Brian} \lastname{Gravelle}\inst{5} %\fnsep\thanks{\email{gravelle@cs.uoregon.edu}}	
  \and \firstname{Matti} \lastname{Kortelainen}\inst{2} %\fnsep\thanks{\email{matti.kortelainen@cern.ch}}
  \and \firstname{Vyacheslav} \lastname{Krutelyov}\inst{4} %\fnsep\thanks{\email{vyacheslav.krutelyov@cern.ch}}
  \and \firstname{Steven} \lastname{Lantz}\inst{1} %\fnsep\thanks{\email{steve.lantz@cornell.edu}}	
  \and \firstname{Mario} \lastname{Masciovecchio}\inst{4} %\fnsep\thanks{\email{mario.masciovecchio@cern.ch}}
  \and \firstname{Kevin} \lastname{McDermott}\inst{1} %\fnsep\thanks{\email{kevin.mcdermott@cern.ch}}	
  \and \firstname{Boyana} \lastname{Norris}\inst{5} %\fnsep\thanks{\email{norris@cs.uoregon.edu}}	
  \and \firstname{Michael} \lastname{Reid}\inst{1}   
  \and \firstname{Allison} \lastname{Reinsvold Hall}\inst{2} %\fnsep\thanks{\email{ahall@fnal.gov}}
  \and \firstname{Daniel} \lastname{Riley}\inst{1} %\fnsep\thanks{\email{daniel.riley@cornell.edu}}	
  \and \firstname{Matev\v{z}} \lastname{Tadel}\inst{4}
  %\fnsep\thanks{\email{mtadel@ucsd.edu}}	
  \and \firstname{Peter} \lastname{Wittich}\inst{1} %\fnsep\thanks{\email{wittich@cornell.edu}}	
  \and \firstname{Bei} \lastname{Wang}\inst{3} 
  \and \firstname{Frank} \lastname{W\"{u}rthwein}\inst{4} %\fnsep\thanks{\email{fkw@ucsd.edu}}	
  \and \firstname{Avraham} \lastname{Yagil}\inst{4} %\fnsep\thanks{\email{ayagil@physics.ucsd.edu}}	
}

\institute{Cornell University, Ithaca, NY, USA 14853
  \and     Fermi National Accelerator Laboratory, Batavia, IL, USA 60510
  \and     Princeton University, Princeton, NJ, USA 08544
  \and     UC San Diego, La Jolla, CA, USA 92093
  \and     University of Oregon, Eugene, OR, USA 97403
}
\abstract{%
  One of the most computationally challenging problems expected for the High-Luminosity Large Hadron Collider (HL-LHC) is finding and fitting particle tracks during event reconstruction. Algorithms used at the LHC today rely on Kalman filtering, which builds physical trajectories incrementally while incorporating material effects and error estimation. Recognizing the need for faster computational throughput, we have adapted Kalman-filter-based methods for highly parallel, many-core SIMD and SIMT architectures that are now prevalent in high-performance hardware. Previously we observed significant parallel speedups, with physics performance comparable to CMS standard tracking, on Intel Xeon, Intel Xeon Phi, and (to a limited extent) NVIDIA GPUs. While early tests were based on artificial events occurring inside an idealized barrel detector, we showed subsequently that our \mkfit software builds tracks successfully from complex simulated events (including detector pileup) occurring inside a geometrically accurate representation of the CMS-2017 tracker. Here, we report on advances in both the computational and physics performance of \mkfit, as well as progress toward integration with CMS production software. Recently we have improved the overall efficiency of the algorithm by preserving short track candidates at a relatively early stage rather than attempting to extend them over many layers. Moreover, \mkfit formerly produced an excess of duplicate tracks; these are now explicitly removed in an additional processing step. We demonstrate that with these enhancements, \mkfit becomes a suitable choice for the first iteration of CMS tracking, and eventually for later iterations as well. We plan to test this capability in the CMS High Level Trigger during Run 3 of the LHC, with an ultimate goal of using it in both the CMS HLT and offline reconstruction for the HL-LHC CMS tracker.
}
\maketitle
\section{Introduction}
\label{intro}

The reconstruction of charged particle tracks (tracking) is crucial for the physics goals of the Large Hadron Collider (LHC) experiments as it is needed to estimate the particle momenta, identify the particle type, tag the flavor of hadron jets, and improve the resolution of both jet energy and missing transverse momentum. Tracking is typically the most time consuming reconstruction task and its time scales poorly with the detector occupancy; this is a problem as the LHC luminosity increases and leads to a larger and larger number of proton interactions per beam crossing (pile-up, or PU). The challenge will be even greater at the High Luminosity LHC and especially at the High Level Trigger (HLT), an accurate measurement is needed quickly so that the most interesting events are stored for further processing. The time budget constraint needs to be addressed either by reducing the tracking phase space, with clear negative consequences on the experiments' physics reach, or by speeding up the tracking algorithms.

In the last 10-15 years, the continuous exponential increase in the number of transistors per processor did not lead anymore to an exponential increase in clock frequency; therefore the needed speedup will not come ``for free'' and will require a substantial rework of the algorithms. For modern processors, most of the overall performance increase (in terms of overall floating point operations per second) comes from parallelizing over a large number of logical cores, while more moderate single-thread performance improvements are due to vector or SIMD (single instruction multiple data) operations. In light of these trends, algorithms should be reworked such that paralellism at both the data and instruction levels can be exploited.

The mission of the \mkfit project\footnote{http://trackreco.github.io/} is to speed up Kalman filter (KF) tracking algorithms~\cite{Fruhwirth} using highly parallel architectures. KF is widely used for tracking because the high efficiency of its physics performance is well understood and it includes a robust handling of material effects. The KF is a two-step iterative process: the track state (parameters and their uncertainties) are propagated from layer N-1 to layer N; then, the track state is updated using the position estimate provided by a detector measurement (\emph{hit}). The two KF-based steps of tracking are track building and track fitting. Track building is a combinatorial search of the hits belonging to a given track; typically it is initiated from a proto-track (\emph{seed}) defined in the innermost detector layers, and proceeds outwards. Fitting simply applies the KF procedure to all hits identified during building in order to extract the best estimate of the track parameters. Track building takes the largest fraction of time and it is not straightforward to parallelize efficiently due to the branching required by the combinatorial exploration of many possible track candidates, and due to the work imbalance caused by the variability inherent in the data (different number of hits per track and tracks per event). Further computing challenges are posed by the many operations with small-size matrices leading to low arithmetic intensity, and the large number of hits per event (order of 100k at PU=70).

\section{Algorithm Overview}

Our implementation of the KF tracking algorithm is described in detail elsewhere~\cite{pkf-fit,pkf-finding,pkf-ahmdal,pkf-clone-engine,pkf-tbb,pkf-gpu,pkf-cms-geom,pkf-geom-plugin,pkf-short-trks,pkf-duplicates} and it has not fundamentally changed, so only its main features will be summarized here.

KF operations are implemented using the Matriplex library\footnote{https://github.com/trackreco/mkFit/tree/devel/Matriplex}, a matrix-major representation that allows for SIMD processing of track candidates;
Matriplex auto-generates vectorized code that is aware of the matrix sparsity. The algorithm is multithreaded using \tbb tasks, where independent tasks are created at multiple levels: for different events, different detector regions, and for different seeds (grouped in bunches). Due to the limited bandwidth and cache size of parallel architectures, we implemented a lightweight description of detector in terms of geometry, material, magnetic field; for instance, the geometry representation collapses barrel (forward) layers at average r (z), and has no knowledge of detector modules so it relies on the 3D position of hits. Within the combinatorial part of the algorithm we minimize memory operations both in their number and in their size: this requires compact data formats and a bookkeeping of the explored track candidates, so that only the best ranking ones are cloned at each layer (with a per-seed cap).

\begin{figure*}
\centering
\includegraphics[width=6.4cm,clip]{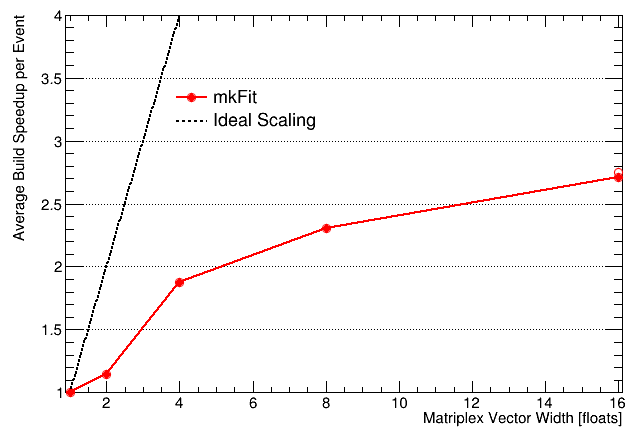}
\hfill
\includegraphics[width=6.4cm,clip]{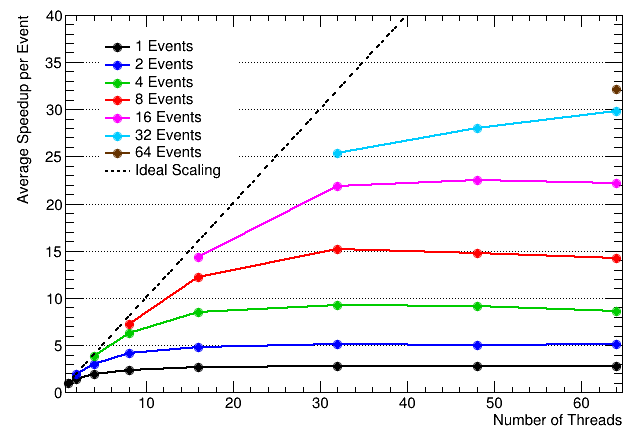}
\caption{\mkfit speedups on Skylake. Left: speedup of the single-threaded track building core algorithm as a function of the vector width configured in the Matriplex library; the open circle visible at a width of 16 floats corresponds to the usage of vector intrinsics code in Matriplex. Right: speedup of the full track building application as a function of the total number of threads (for vector width 16); different curves correspond to a different number of threads dedicated to event-level tasks.}
\label{fig-standalone}
\end{figure*}

The standalone implementation of our algorithm has been tested on a 32-core system with dual Intel Xeon Gold 6130 processors (Skylake, \emph{SKL}), see Fig.~\ref{fig-standalone}. Hyperthreading is enabled, but Turbo Boost is disabled to emphasize the scaling behavior of the code over that of the hardware. For Gold 6130, this means the clock frequency is fixed at 2.1 GHz for all cores, though it drops by 10\% if more than 11 threads are active per processor. The core of the track building algorithm achieves nearly 3x speedup from vectorization when using a vector size of 16 floats. Ahmdal’s law implies that 60-70\% of core algorithm code is effectively vectorized. We note that the full track building application includes data preparation and cleaning steps for each event, and these steps are not vectorizable. Using multithreading, we achieve further speedups of more than a factor of 30 compared to the single-threaded, vectorized execution of the full application. The multithreaded scaling is close to ideal when the number of threads does not exceed the number of available physical cores (32), and all threads are dedicated to different events (which is equivalent to a multi-process execution over multiple events).

\section{Integration in \cmssw}

The deployment of our code in the framework of the CMS experiment, \cmssw, has recently been the main focus of the group. \mkfit is currently integrated in \cmssw as an external package, pulled from a public repository, compiled, and distributed along with \cmssw releases. This is a major achievement that makes our work available to the whole collaboration. Two aspects are not ideal in this first integration: first, within a central \cmssw release \mkfit is compiled with \gcc using the core2 instruction set (limiting vectors to 128 bits), and second, dedicated steps are used to convert \cmssw data formats to and from the \mkfit ones.

The CMS tracking~\cite{CMS-tracking} is structured in more than ten iterations, where each iteration performs the same steps: track seeding, building, fitting, and selection. Hits used in tracks reconstructed with high quality are masked to reduce the combinatorics of the next iteration. As a first milestone, our focus has been the track building for the initial iteration, based on seeds from the four innermost pixel layers and targeting prompt tracks. However, there is nothing specific to the first iteration, so the application of \mkfit could be extended to other iterations as well.

\begin{figure*}
\centering
\includegraphics[width=6.4cm,clip]{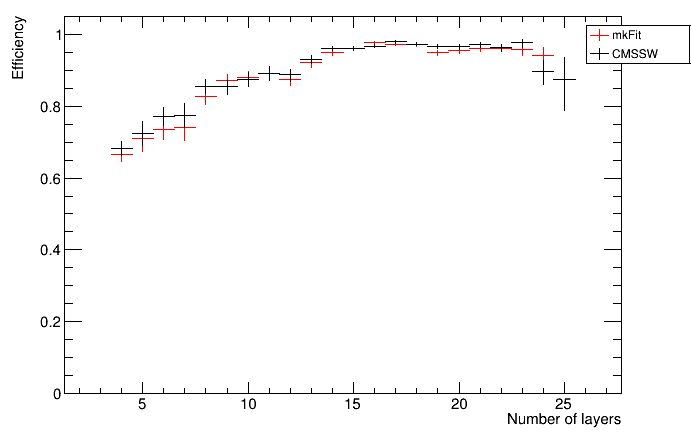}
\hfill
\includegraphics[width=6.4cm,clip]{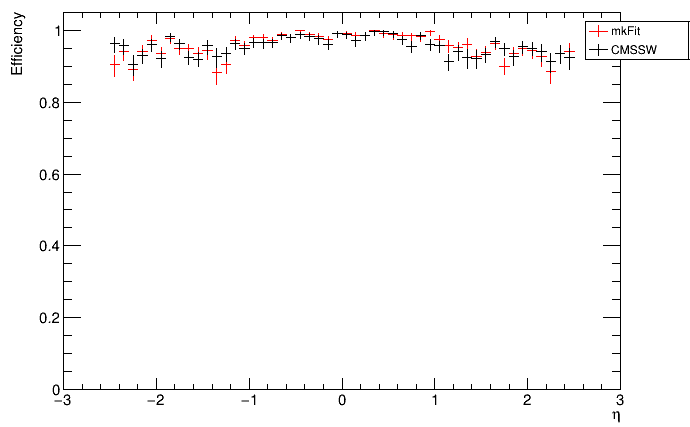}
\caption{Efficiency for \mkfit and \cmssw algorithms as a function of the number of crossed layers and the pseudorapidity $\eta$ of the simulated particle, for simulated particles with $p_{\rm T}$$>$0 and $p_{\rm T}$$>$0.9 GeV, respectively. The sample used are $t\bar{t}$ events with an average PU of 50. The track building efficiency is defined as the ratio of the number of simulated tracks that can be matched to a reconstructed track to the number of simulated particle tracks that can be matched to an initial iteration seed. The matching criterion is that more than 75\% of the reconstructed track or seed hits originates from the simulated track.}
\label{fig-eff}
\end{figure*}

The integration in \cmssw gave access to the central validation tools, which revealed a subpopulation in the initial iteration where the \mkfit efficiency was suffering: short tracks. After updating the logic to count the number of missing hits in a track in a consistent way, and after re-tuning the score used to decide which candidates are the best ones, we recovered efficiency for short tracks so that our efficiency is on par with \cmssw across the board (Fig.~\ref{fig-eff}). Although the \mkfit performance in terms of fake and duplicate rate is currently at an acceptable level, some more work is still needed to bring them down to the \cmssw level. Another feature that we plan to implement in \mkfit is the recovery of hits in overlapping detector modules, so that more than one hit per layer can be assigned to tracks.

\begin{figure}[h]
\centering
\includegraphics[width=6cm,clip]{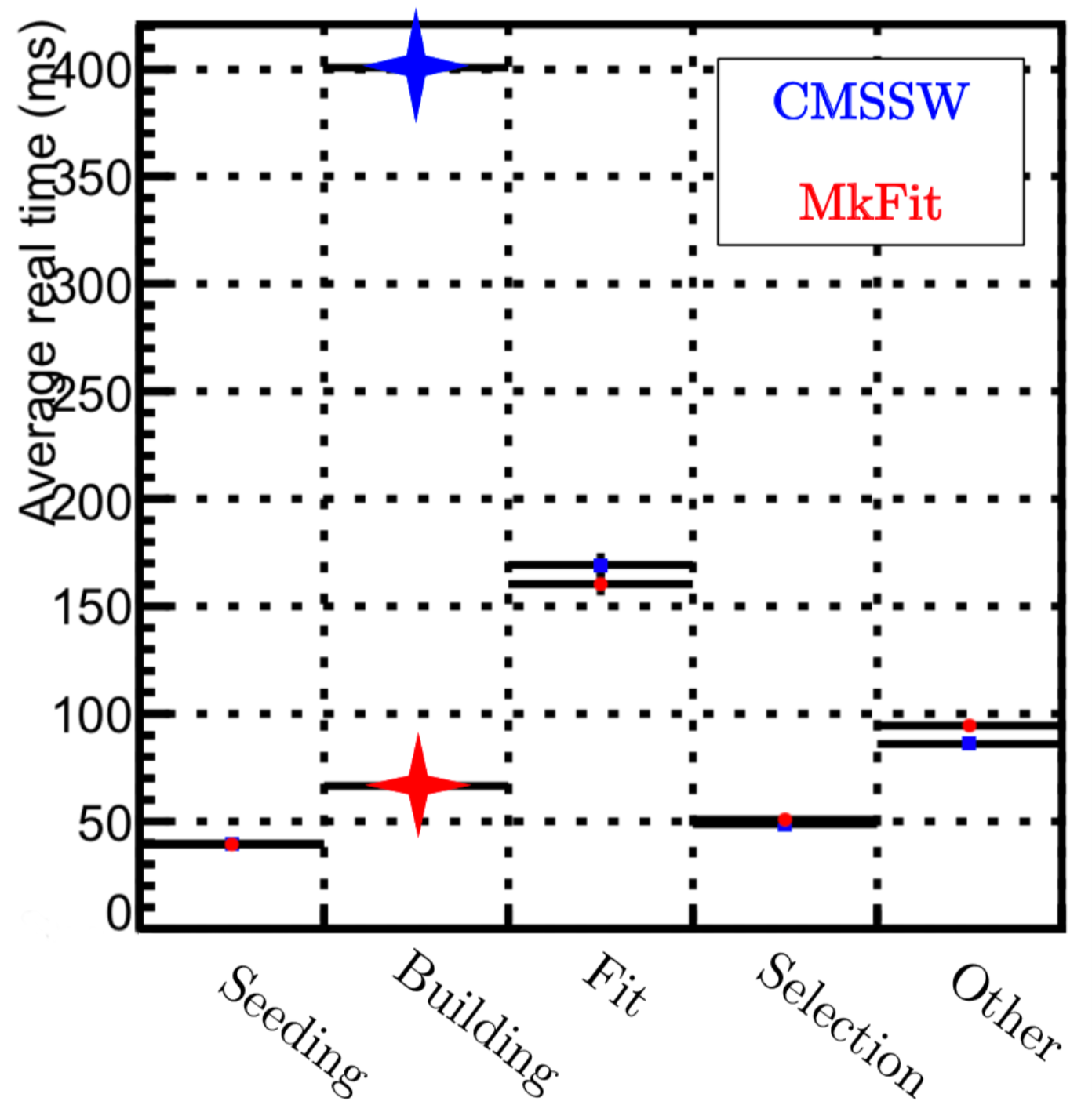}
\caption{Time for the various steps of the initial iteration for a single-threaded execution of the \cmssw tracking on SKL. Track building with \mkfit is compared to the \cmssw default. The sample used are $t\bar{t}$ events with an average PU of 50.}
\label{fig-cmssw-time}
\end{figure}

The initial iteration time for a single-threaded execution of the \cmssw tracking on SKL has been evaluated using the \cmssw default track building as well as \mkfit (Fig.~\ref{fig-cmssw-time}); the sample used consists of $t\bar{t}$ events with an average PU of 50. We observe a speedup of a factor 6.2 when using \mkfit. It is interesting to note that when \mkfit is used the track building step becomes faster than the track fit. Data format conversions between \cmssw and \mkfit account for about 25\% of \mkfit time, so a larger speedup is possible in case the data formats are harmonized in a way that allows for the conversion step to be reduced or entirely removed. In this test \mkfit is compiled with the Intel compiler \icc 19.0.4.243 enabling the AVX-512 instruction set; if we were to use the default \cmssw compilation settings (\gcc 7.3.1, core2) the speedup would reduce by more than a factor of two.

\section{Towards an HLT implementation}

While the work so far mostly focused on applying \mkfit to CMS tracking with the offline configuration, the most natural environment for a first deployment of \mkfit is the HLT. The CMS HLT configuration has different challenges with respect to offline: first, in many HLT paths, tracking is run only in regions of interest around the direction of leptons or jets, and second, the silicon strip hit reconstruction is performed on-demand within the track building. On-demand hit reconstruction means that the raw data are not processed upfront for all modules in the silicon strip detector; instead, when a track candidate first searches for hits on a given module, the module raw data are processed and then cached for subsequent access. The on-demand option is currently not supported within \mkfit, so all hits need to be available upfront. With the tools currently available in \cmssw, the global strip hit reconstruction chain is costly, so we need to investigate a faster implementation if we want to profit from the \mkfit speedups at HLT. Our goal is to start from raw strip detector data and output hits in the \mkfit data format; we also require that the new implementation be compatible with GPU execution. The steps to be performed in the strip processing are: raw data unpacking and remapping, strip data calibration, strip data clustering, and evaluation of the 3D hit position. Here we report on a preliminary implementation of the clustering algorithm.

In its current version the algorithm mimics the \cmssw implementation and works as follows: first, seed strips with ADC values greater than 3 times the expected noise are identified; second, clusters are formed adding to the seed all strips with ADC values above twice the expected noise that are either consecutive or have a small gap (the size of the allowed gap depends on the presence of bad strips in the module); finally, the cluster is required to have a quadrature sum of the strip ADCs larger than 5 times the quadrature sum of the expected noise, and to have a minimum total charge. We developed a first implementation both for CPU (C++ code parallelized using \texttt{OpenMP}) and for GPU (using \texttt{CUDA}). Preliminary tests on single events show that overheads currently account for the dominant fraction ($>$90\%) of the time for the GPU execution, where overheads include data transfer and memory allocation. Work is ongoing towards an improved implementation that reduces the overheads by processing multiple events concurrently, thus allowing for asynchronous memory transfer, and by using a memory pool to pay the allocation overhead only at the beginning and end of job.

\section{Studies Towards a Portable Implementation}

Given that the processor market, HPC centers, and the experiments' online farms are all trending towards an increase in the utilization of GPUs, the exploration of GPU-compatible versions of our algorithm is becoming a priority. A portable solution would guarantee that the code is maintainable, with minimal differences between CPU and GPU versions, even if such portability may require trade-offs in terms of performance.

We started a collaboration with the RAPIDS\footnote{SciDAC Institute for Resource and Application Productivity through computation, Information, and Data Science. SciDAC is the Scientific Discovery through Advanced Computing program.} group at Oak Ridge National Laboratory to explore usage of compiler directives for portability between CPU and GPU. First developments in this respect include a version of \mkfit where threads are managed with \texttt{OpenMP} (working only on CPU for now), and testing \texttt{OpenACC} for a single function. There is not a dominant function, taking most of the compute time in our code, so the choice of which function to study was somewhat arbitrary. We chose the "PropagationToZ" function as it is one of the top contributors but it is simple enough to be an ideal candidate for exploration tests. We compared a serial version with various parallel versions, and found an \texttt{OpenMP} CPU version to be 23.9 times faster than the serial, and an \texttt{OpenACC} version to be 207.1 (375.1) times faster than the serial CPU version on a single GPU when accounting for (excluding) the data transfer. Tests were performed on a Summit supercomputer node (dual IBM POWER9 CPUs, each with 22 physical cores, plus 6 NVIDIA Volta V100 GPUs). While these results are encouraging, there are significant challenges ahead, including the management of data transfers, and the interface with \cmssw.
%; in particular the latter may be an insurmountable problem as CMSSW is currently not planning to support compiler directives
%Other tests towards GPU-compatible code:
%array programming: xtensor/numpy/cupy
%plan to try portable libraries and revisit CUDA implementation

%\begin{figure}[h]
%\centering
%\includegraphics[width=6cm,clip]{figs/p2z-Summit-test.png}
%\label{fig-p2z}
%\end{figure}

\section{Conclusions}

In summary, the \mkfit code has been integrated in \cmssw as an external library, and shows physics performance in terms of efficiency that is on par with the current \cmssw tracking algorithm, while running more than 6 times faster when \mkfit is compiled with \icc and AVX-512. The utilization of GPUs is being explored at different stages, including strip hit reconstruction for the HLT application and elements of a portable implementation of the track building algorithm.

\section{Acknowledgements}

This work is supported by the U.S. National Science Foundation, under the grants PHY-1520969, PHY-1521042, PHY-1520942, PHY-1624356, and OAC-1836650, and by the U.S. Department of Energy, Office of Science, Office of Advanced Scientific Computing Research and Office of High Energy Physics, Scientific Discovery through Advanced Computing (SciDAC) program.
%under Award Number(s) XXXXXXX.

%
% BibTeX or Biber users please use (the style is already called in the class, ensure that the "woc.bst" style is in your local directory)
% \bibliography{name or your bibliography database}

\begin{thebibliography}{}

\bibitem{Fruhwirth} Rudolf Fr{\"u}hwirth, Nucl. Instrum. Meth. \textbf{A262}, 440--450 (1987)

\bibitem{CMS-tracking} S.~Chatrchyan {\it et al.} [CMS Collaboration], JINST {\bf 9}, no. 10, P10009 (2014)

\bibitem{pkf-fit} Giuseppe Cerati \etal, J. Phys.: Conf. Ser. \textbf{608}, 012057 (2015) %ACAT2015

\bibitem{pkf-finding} Giuseppe Cerati \etal, J. Phys.: Conf. Ser. \textbf{664}, 072008 (2015) %CHEP2015

\bibitem{pkf-ahmdal} Giuseppe Cerati \etal, doi:10.1109/NSSMIC.2015.7581932 %IEEE2015

\bibitem{pkf-clone-engine} Giuseppe Cerati \etal, EPJ Web of Conferences \textbf{127}, 00010 (2016) %CTD2016

\bibitem{pkf-tbb} Giuseppe Cerati \etal, J. Phys.: Conf. Ser. \textbf{898}, 042051 (2017) %CHEP2016

\bibitem{pkf-gpu} Giuseppe Cerati \etal, EPJ Web of Conferences \textbf{150}, 00006 (2017) %CTD2017

\bibitem{pkf-cms-geom} Giuseppe Cerati \etal, J. Phys.: Conf. Ser. \textbf{1085}, 042016 (2018) %ACAT2017

\bibitem{pkf-geom-plugin} Giuseppe Cerati \etal, EPJ Web Conf.\  {\bf 214}, 02002 (2019) %CHEP2018

\bibitem{pkf-short-trks} Giuseppe Cerati \etal, arXiv:1906.02253 [physics.ins-det] %ACAT2019

\bibitem{pkf-duplicates} Giuseppe Cerati \etal, arXiv:1906.11744 [physics.ins-det] %CTD2019
  
\end{thebibliography}

\end{document}